\documentclass[a4paper,accepted=2025-03-11]{quantumarticle}
\pdfoutput=1

\usepackage{amsfonts,amsmath,amssymb}

\usepackage[dvipsnames]{xcolor}
\usepackage{tikz}
\usetikzlibrary{arrows.meta, calc}
\usepackage{physics}
\usepackage{graphicx}
\usepackage{dcolumn}
\usepackage{bm}
\usepackage[breaklinks]{hyperref}
\hypersetup{colorlinks=true, linkcolor=blue, citecolor=blue, filecolor=blue, urlcolor=blue}
\usepackage[capitalise]{cleveref}
\usepackage{ulem}
\usepackage{soul}
\usepackage{mathtools}
\usepackage{cancel}
\usepackage{tabularray}

\usepackage[sort,numbers]{natbib}

\newcommand{\bigO}[1]{\mathcal{O}\qty(#1)}

\tikzset{->/.style={-{Stealth[]}}}
\tikzset{<-/.style={{Stealth[]}-}}

\NewDocumentCommand{\fswp}{m}{
	\draw #1 node {$\boldsymbol{\times}$} -- +(1,0) node {$\boldsymbol{\times}$};
}

\NewDocumentCommand{\Ugates}{mO{1}}{
	\foreach\x in #1
	{
		\draw[xshift=\x cm,fill=white] (-0.25,#2) rectangle +(1.5,0.75);
	}
}

\NewDocumentCommand{\UgateLR}{O{1}}{
	
	\begin{scope}
		\def\yRec{0.75}

		\draw[fill=white] (0.75,#1) rectangle +(0.5,\yRec);
		\draw[xshift=7cm, fill=white] (0.75,#1) rectangle +(0.5,\yRec);

		\draw[color=white] (0.75,{#1 - 0.05}) -- +(0,{\yRec+0.11});
		\draw[xshift=7cm, color=white] (1.25,{#1 - 0.05}) -- +(0,{\yRec+0.11});
	\end{scope}
}

\begin{document}

\title{Free fermions under adaptive quantum dynamics}

\author{Vikram Ravindranath}
\email{vikram.ravindranath@bc.edu}
\affiliation{Department of Physics, Boston College, Chestnut Hill, MA 02467, USA}

\author{Zhi-Cheng Yang}
\email{zcyang19@pku.edu.cn}
\affiliation{School of Physics, Peking University, Beijing 100871, China}
\affiliation{Center for High Energy Physics, Peking University, Beijing 100871, China}

\author{Xiao Chen}
\email{xiao.chen@bc.edu}
\affiliation{Department of Physics, Boston College, Chestnut Hill, MA 02467, USA}

\date{March 17, 2025}

\begin{abstract}

We study free fermion systems under adaptive quantum dynamics consisting of unitary gates and projective measurements followed by corrective unitary operations. We further introduce a classical flag for each site, allowing for an active or inactive status which determines whether or not the unitary gates are allowed to apply. In this dynamics, the individual quantum trajectories exhibit a measurement-induced entanglement transition from critical to area-law scaling above a critical measurement rate, similar to previously studied models of free fermions under continuous monitoring. Furthermore, we find that the corrective unitary operations can steer the system into a state characterized by charge-density-wave order. Consequently, an additional phase transition occurs, which can be observed at both the level of the quantum trajectory and the quantum channel.  We establish that the entanglement transition and the steering transition are fundamentally distinct. The latter transition belongs to the parity-conserving (PC) universality class, arising from the interplay between the inherent fermionic parity and classical labelling. We demonstrate both the entanglement and the steering transitions via efficient numerical simulations of free fermion systems, which confirm the PC universality class of the latter.

\end{abstract}

\maketitle

\section{Introduction}

The research frontier of non-equilibrium quantum many-body dynamics has undergone significant expansion in recent years. This growth can be attributed to the emergence of numerous quantum platforms, which offer an increasing number of qubits and a higher degree of controllability in experimental settings. These platforms enable the simulation of various types of dynamics, ranging from unitary evolution governed by time-independent Hamiltonians to quantum circuit evolutions.

In the realm of quantum circuit dynamics, the unitary gate offers customization options for a multitude of computational tasks, surpassing the capabilities of classical algorithms. Furthermore, the introduction of randomness into the unitary gate has proven to be an invaluable theoretical tool in the study of quantum dynamics~\cite{fisher2023random}, including quantum chaos~\cite{PhysRevLett.123.210603, PhysRevX.8.041019, PhysRevLett.121.060601}, operator spreading~\cite{PhysRevX.8.021014, PhysRevX.8.021013, PhysRevResearch.2.033032, PhysRevX.8.031057}, entanglement growth~\cite{PhysRevX.8.021013, PhysRevX.7.031016, PhysRevB.99.174205, PhysRevX.10.031066}, information scrambling~\cite{mi2021information}, and anomalous transport~\cite{PhysRevLett.127.230602}.

A more exciting direction is the possibility of including mid-circuit measurements into the evolution, which renders the dynamics non-unitary. Despite the similarity to open quantum systems where the coupling of the system to the environment leads to non-unitarity, a key distinction here is that, unlike the environment, experimentalists often have full knowledge of the particular measurement type and location on the system, as well as the measurement record. This makes it reasonable to investigate non-unitary dynamics at the level of individual quantum trajectories, which has led to the identification of a measurement-induced entanglement transition in monitored quantum circuits~\cite{PhysRevB.100.134306, PhysRevB.98.205136, PhysRevB.99.224307, PhysRevX.9.031009}. Such an entanglement transition in turn is only visible upon unravelling the individual quantum trajectories, which poses challenges on its experimental observation~\cite{noel2022measurement, koh2022experimental, hoke2023quantum}.

On the other hand, experimentalists can do a lot more than simply performing measurements and recording the outcomes with these quantum platforms. A key ingredient of \textit{interactive quantum dynamics} is that operations at a later time are determined by the measurement records at earlier times, which necessitates an active dialog between a classical experimentalist and the quantum system. This idea traces back to quantum teleportation, quantum error-correction, and measurement-based quantum computation~\cite{PhysRevLett.86.5188, PhysRevA.68.022312}, and has been further extended to preparation of certain quantum states exhibiting either topological order~\cite{tantivasadakarn2021long, PRXQuantum.3.040337, iqbal2023topological} or continuous symmetry breaking~\cite{hauser2023continuous}.
More recently, aspects of non-equilibrium \textit{dynamical} properties of interactive quantum circuits have started to gain attention~\cite{buchhold2022revealing, iadecola2022dynamical, ravindranath2022entanglement, o2022entanglement, PhysRevLett.130.120402, piroli2022triviality,friedman2022measurement}. Via an adaptive feedback mechanism whereby a corrective unitary operation may be applied following each measurement depending on the outcome, the system is steered towards a particular target state when the feedback rate exceeds a certain threshold. The non-equilibrium universality class of this absorbing-state transition and its interplay with the measurement-induced entanglement transition are both interesting questions that are being actively investigated.

\begin{figure}[thb]
    \centering
    \includegraphics[width=0.5\textwidth]{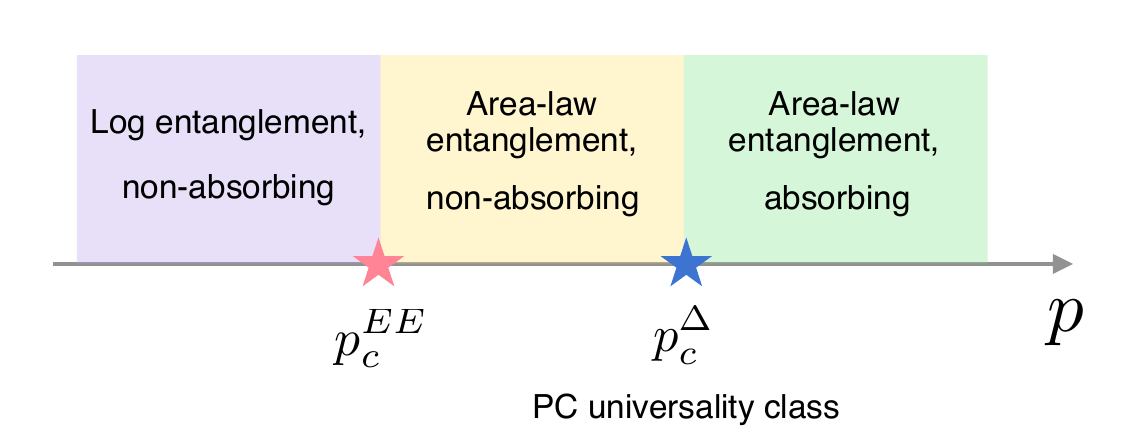}
    \caption{ A schematic phase diagram of free fermions under adaptive quantum dynamics with a fixed feedback rate $r$.}
    \label{fig:phase}
\end{figure}

In this work, we study free fermion systems under adaptive quantum dynamics consisting of unitary evolution and projective measurements followed by corrective unitary operations. The individual quantum trajectories exhibit a measurement-induced entanglement transition from critical to area-law scaling above a critical measurement rate, similar to previously studied models of free fermions under continuous monitoring~\cite{Alberton_2021, PhysRevX.11.041004,Turkeshi_2021,Turkeshi_2022}. In addition, with the assistance of an additional classical register, the combined dynamics of our quantum channel has a nontrivial fixed point given by the N\'eel state $|1010\cdots 10\rangle$. Thus, the trajectory-averaged density matrix further experiences an absorbing-state transition when the feedback rate exceeds a certain threshold. Such a transition can be diagnosed with observables either in the classical register or the actual quantum system. 

The fermion parity symmetry inherent in fermionic systems naturally furnishes a parity conservation in the particle occupation basis, and we find numerically that the absorbing-state transition belongs to the parity-conserving (PC) universality class~\cite{hinrichsen2000non}. Interestingly, unlike previous investigations~\cite{ravindranath2022entanglement,o2022entanglement,piroli2022triviality,PhysRevLett.130.120402}, inferring that the dynamical phase transition belongs to the PC universality class cannot be easily achieved solely based on a feedback mechanism~\cite{ravindranath2022entanglement,o2022entanglement} or the presence of a classical flag~\cite{piroli2022triviality,PhysRevLett.130.120402}. Instead, it arises intrinsically from the combined effect of fermionic parity in the quantum dynamics and the classical labelling. We also demonstrate that the entanglement and absorbing transitions are well-separated in our model (See \cref{fig:phase,fig:PhaseDiag}).

\section{Model}
\label{sec:model}

Our model consists of a one-dimensional, periodic chain of $L$ sites, with each site hosting either 0 or 1 spinless fermion. Additionally, to each site $i$, we associate a
classical ``flag" $f_i$ which takes values 0 (equivalently, labeling the site as ``inactive") or 1 (labeled ``active"); a unitary gate is applied between two sites $i$ and $j$ only if at least one of the sites is ``active" (i.e. only if $f_i + f_j \neq 0$). Sites are labelled inactive -- their flags are set to 0 -- if the state is found to be locally similar to the target state. This local similarity is then prevented from being disrupted by subsequent unitary gates. 

In the unitary layer, we apply random unitary gates to neighboring sites. For concreteness, each gate is independently drawn with equal probability from $\qty{U_1, U_2}$, which are defined as
\begin{equation}
	\begin{aligned}
		U_\alpha \left(i,j\right) = e^{-i\frac{\pi}{4} H_\alpha (i,j)};\\
		H_1 (i,j) = c^\dagger_i c_j + c^\dagger_j c_i,\\
		H_2 (i,j) = c^\dagger_i c^\dagger_j + c_j c_i.
	\end{aligned}
	\label{eq:model}
\end{equation}
$\qty{c^\dagger_i, c_i}$ denote the fermionic creation and annihilation operators on site $i$. However, as we explain in \cref{sec:C_Map}, our results extend to far less restricted gate sets.

\begin{figure*}[tbh]
    \raggedright
    \begin{tikzpicture}[baseline=12em]

		\draw[->,line width=2pt] (0,1) -- node [label={[font=\Large]left:$t$}]{} +(0,5);
		
		\draw[->,line width=2pt,shift={(1,-1)}] (1.2,0) -- node [label={[font=\Large]below:$x$}]{} +(5,0);
		
		\foreach \x in {1,...,8}{
			\draw[xshift=\x cm, line width=1pt] (0,0) -- +(0,8);}
			
		\begin{scope}[
			color=Goldenrod,
			line width=2pt]
			
			\draw (1,1.8) -- (1,2.5);
			\draw (2,1.8) -- (2,2.5);
			\draw (1,5.8) -- (1,6.5);
			\foreach\x in {2,3,4}{
				\draw (\x,5.8) -- (\x,8);
			}
			\draw (4,3.8) -- (4,4.5);
			\draw (5,3.8) -- (5,8);
			\draw (6,3.8) -- (6,6.5);
			\draw (7,3.8) -- (7,4.5);
		\end{scope}
			
		\Ugates{{1,3,...,8}}[0.5];
		\def\ysepC{2.5}
		\Ugates{{2,4,6}}[\ysepC];
		\UgateLR[\ysepC]
		
		\Ugates{{1,3,7}}[4.5]
		\def\ysepC{6.5}
		\Ugates{{6}}[\ysepC]
		\UgateLR[\ysepC]

		\begin{scope}[
			every node/.style={draw=black,fill=red,circle,inner sep=2pt}]
			\foreach \x in {1,...,4}{
				\node at (\x,1.5) {};
			}
			
			\foreach \x in {1,4,5,...,8}{
				\node at (\x,3.5) {};
			}
			
			\foreach \x in {1,...,6}{
				\node at (\x,5.5) {};
			}
		\end{scope}
		
		\begin{scope}[
			every node/.style={fill=white,inner xsep=0,inner ysep=1},
			color=green!50!black]
			
			\fswp{(1,1.8)}
			\fswp{(4,3.8)}
			\fswp{(6,3.8)}
			\fswp{(1,5.8)}
			\fswp{(3,5.8)}
			
		\end{scope}			
	\end{tikzpicture}\hspace{5em}
	\begin{tblr}{colspec={Q[c]Q[l]},rows={3em},stretch=0}
		\tikz[baseline=0.5em]\draw[line width=1pt] (0,0) -- +(0,0.6); & {Active site} \\
		\tikz[baseline=0.5em]\draw[color=Goldenrod,line width=2pt] (0,0) -- +(0,0.6); & {Inactive site}\\
		\tikz[baseline=0.3em]\draw (0,0) rectangle (0.75,0.5); & {$U_1$ or $U_2$}\\
		\tikz[baseline=-0.3em,color=green!50!black,every node/.style={fill=white,inner xsep=0pt, inner ysep=1pt}] \fswp{(0,0)}; & Swap\\
		\tikz[every node/.style={circle, draw=black, fill=red,inner sep=2pt}] \node at (0,0) {}; & $\hat{n}$ Measurement\\
	\end{tblr}
    \caption{A realization of the circuit model considered in this work, with periodic boundary conditions. $U_1$ and $U_2$ [see Eq.~(\ref{eq:model})] are randomly applied with equal probability on two neighboring sites, if at least one of the sites is active. Projective measurements of $\hat{n}_j$ and $\hat{n}_{j+1}$ are always performed in pairs. If the outcome is $(n_{2j-1}, n_{2j})=(0, 1)$ in the odd layer, or $(n_{2j}, n_{2j+1})=(1,0)$ in the even layer, a SWAP gate is applied on the two measured sites with feedback rate $r$.
    The pair of measured sites is subsequently ``deactivated", unless the measurement outcome is $(1,1)$ or $(0,0)$.}
    \label{fig:circuit}
\end{figure*}
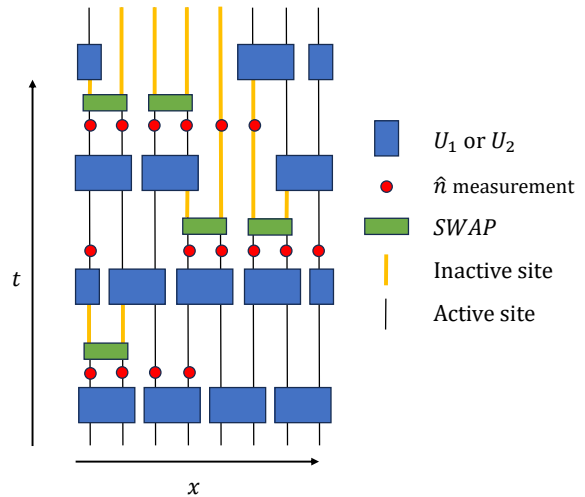

The complete protocol, depicted in \cref{fig:circuit}, is designed to steer initial states towards a ``target" state $\ket{\psi_{\rm targ}} \equiv \ket{10101010\dots}$. Each time-step entails the following:
 \begin{enumerate}
 \item{Initialization: start from an initial product state in the $\{n_i\}$ basis $|\psi_0\rangle$. The classical flag is initialized to be ``active" on all sites, i.e. $f_i=1$, $\forall \ i$;}

 \item{Unitary evolution on \textit{odd} links: randomly apply $U_1$ or $U_2$ with equal probability on bonds $(2j-1, 2j)$, if $f_{2j-1}=1$ or $f_{2j}=1$. Set both sites as ``active" after applying the unitary gate: $f_{2j-1}=f_{2j}=1$;}

 \item{Measurement on \textit{odd} links: with probability $p$, measure the occupation numbers on two neighboring sites connected by an odd link $(n_{2j-1}, n_{2j})$}:
 \begin{itemize}
    \item if the measurement outcomes are $(1,1)$ or $(0,0)$, do nothing;

    \item if the measurement outcomes are $(1,0)$, set both sites to ``inactive", i.e. $f_{2j-1}=f_{2j}=0$;

    \item if the measurement outcomes are $(0,1)$, with probability $r$, apply a SWAP gate on the two measurement sites, then set both sites to ``inactive" $f_{2j-1}=f_{2j}=0$;
   
 \end{itemize}

  \item{Unitary evolution on \textit{even} links: randomly apply $U_1$ or $U_2$ with equal probability on bonds $(2j, 2j+1)$, if $f_{2j}=1$ or $f_{2j+1}=1$. Set both sites as ``active" after applying the unitary gate: $f_{2j}=f_{2j+1}=1$;}

 \item{Measurement on \textit{even} links: with probability $p$, measure the occupation numbers on two neighboring sites connected by an even link $(n_{2j}, n_{2j+1})$}:
 \begin{itemize}
    \item if the measurement outcomes are $(1,1)$ or $(0,0)$, do nothing;

    \item if the measurement outcomes are $(0,1)$, set both sites to ``inactive", i.e. $f_{2j}=f_{2j+1}=0$;

    \item if the measurement outcomes are $(1,0)$, with probability $r$, apply a SWAP gate on the two measurement sites, then set both sites to ``inactive" $f_{2j}=f_{2j+1}=0$.
   
 \end{itemize}

 \end{enumerate}
 Steps $2-5$ are then repeated at each time. For each individual quantum trajectory, the fermionic parity $P\equiv\prod\limits_{i=1}^L (1-2\hat{n}_i)$ is conserved.

We consider quantum trajectories under the hybrid adaptive dynamics described above. For sufficiently large $p$ and $r$, we expect that all quantum trajectories will be steered towards a target state which has the occupation pattern $\ket{\psi_{\rm targ}} \equiv \ket{10101010\dots}$, regardless of the choice of initial state.

On the other hand, when either $p$ or $r$ is very small, each state will evolve to a superposition of an extensive number of fermionic configurations. This suggests that a dynamical \textit{absorbing} phase transition is induced by active steering. To investigate this phase transition, we propose utilizing two complementary quantities as diagnostic tools, one for the classical flag variable and the other for the quantum system: 
(1) the density of active sites, given by 
\begin{equation}
    \rho_{\rm active} (t) = \frac{1}{L}\sum\limits_{i=1}^L f_i (t),
\end{equation}
and (2) the charge imbalance between neighboring sites
\begin{equation}
	\Delta (t) = \frac{1}{L}\sum\limits_{i=1}^{L}(-1)^{i+1}\Big(\expval{\hat{n}_i} - \expval{\hat{n}_{i+1}}\Big).
\end{equation}

We also study the behavior of the entanglement as the rates of the measurement and feedback are varied. We are interested in the R\'enyi entropies $S^{(n)}_A$ of a subsystem $A$, defined as 
\begin{equation}
    S^{(n)}_A = \frac{1}{1-n} \log\qty(\Tr\rho_A^n),
    \label{eq:renyient}
\end{equation}
where $\rho_A = \Tr_{\bar{A}}\dyad{\psi}$ is the reduced density matrix that describes the subsystem $A$ ($\bar{A}$ is its complement in the whole system). 

Similar to previous studies in free fermions under continuous monitoring~\cite{Alberton_2021,10.21468/SciPostPhys.7.2.024}, we expect to find a measurement-induced phase transition (MIPT) between a regime with logarithmic entanglement scaling (where $S_A\sim \log|A|$), and an area law regime ($S_A \sim \mathcal{O}(1)$), which we verify via numerical simulations.

\section{Numerical Results}

Free fermion systems can be exactly simulated for large system sizes on classical computers. Such fermionic Gaussian states are completely described by a correlation matrix $C$ containing single particle correlations:
\begin{equation}
    C =  
    \begin{pmatrix}
    \langle c_i c^\dagger_j \rangle & \expval{c_i c_j}  \\
    \langle c^\dagger_i c^\dagger_j \rangle & \langle c^\dagger_i c_j \rangle
    \end{pmatrix}.
    \label{eq:Cmat}
\end{equation}
In particular, an expression for $S^{(n)}_A$ can be obtained from $C_A$ -- the restriction of $C$ to fermionic operators in subsystem $A$ -- giving

\begin{equation}
    S^{(n)}_A = \frac{1}{2(1-n)} \Tr \left\{(\log[C^n_A + (1-C_A)^n]\right\}.
\end{equation}

A numerical procedure to simulate the dynamics of a pure state that incorporates measurements without relying on the inversion of this matrix \cite{BravyiFLO} is detailed in \cref{app:numerics}. The gist of this method utilizes the fact that a free fermion state $\ket{\psi}$ defined on $L$ sites can be uniquely determined by the $L$ operators that annihilate it. The correlation matrix can then be straightforwardly obtained from these operators.

The results of numerical simulations of the free fermionic circuit defined above are presented in this section. We first demonstrate an MIPT consistent with previous studies on free fermions \cite{Alberton_2021,PhysRevX.11.041004}. In a departure from these studies, our circuit also exhibits an absorbing phase transition as $p$ and $r$ are varied, where the measurement outcome-averaged late time density matrix is a mixture of extensively many random states on one side of this transition, and (almost) a pure state $\dyad{101010\dots}$ on the other.

\subsection{Measurement-Induced Entanglement Phase Transition}
\label{sec:MIPT}

\begin{figure}
    \centering
    \includegraphics[width=0.475\textwidth]{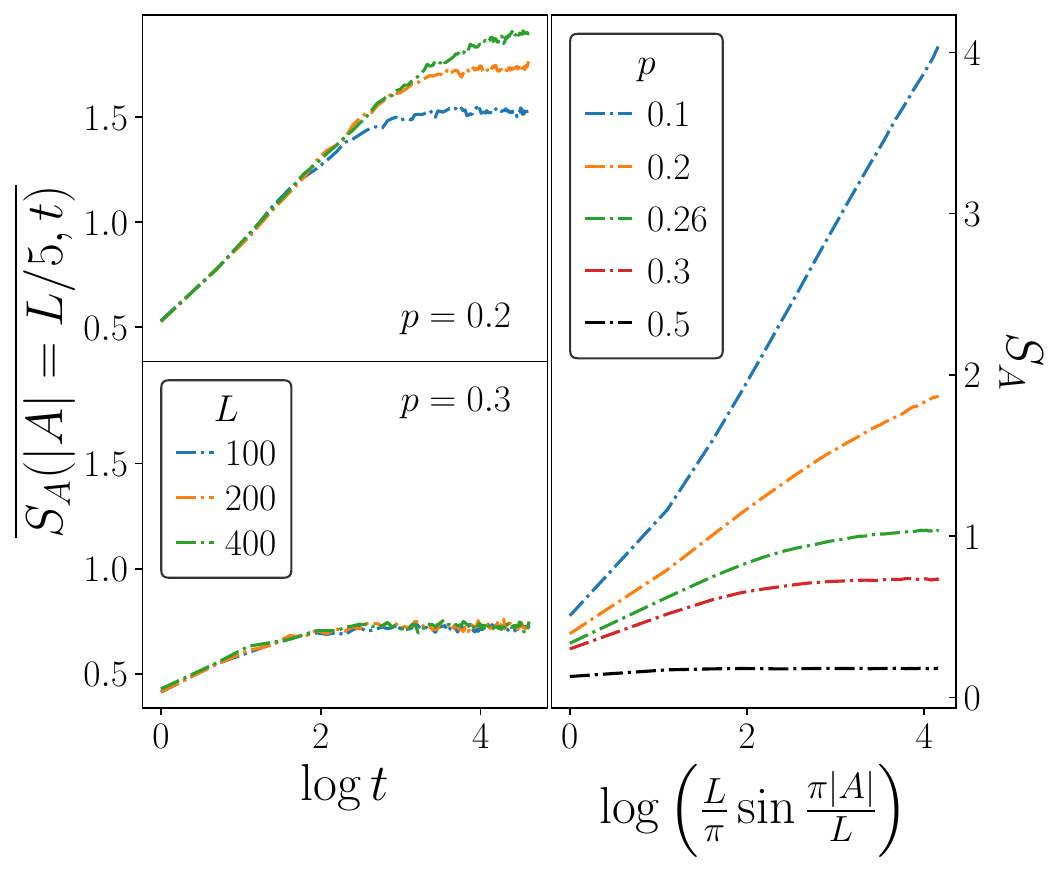}
    \caption{The entanglement entropy for contiguous subsystems $A$, shown for different $p$, at $r=1$, showing a change in both dynamical and steady-state behavior. (Left) $S_A$ is plotted as a function of $\log t$ for different system sizes, for subsystem sizes which are a fixed fraction (0.2) of the size of the whole system $L$. Distinct dynamical behavior is observed, from a logarithmic growth at small $p$, culminating in an $L$-dependent value, to a brief transient growth saturating to an $L$-independent value at large $p$. (Right) Plotted as a function of varying subsystem size $\log\qty(\frac{L}{\pi}\sin \frac{\pi|A|}{L})$ for a system of $L=200$, we see a change from a logarithmic to an area-law scaling with the size of the subsystem at $p^{EE}_c\sim0.26$.}
    \label{fig:ents}
\end{figure}

We begin by fixing the feedback rate $r=1$, and explore how the properties of the system change upon varying the measurement rate $p$. We find two distinct scaling behaviors of the subsystem entanglement entropy $S^{(2)}_A$, as shown in \cref{fig:ents}. For small $p$, we find that a logarithmic scaling form
\begin{equation}
    S_A \sim \alpha \log\qty(\frac{L}{\pi}\sin \frac{\pi|A|}{L})    
\end{equation}
is obeyed, with $\alpha$ being a $p$-dependent parameter. As we increase $p$ beyond a critical value of approximately $p_c \approx 0.26$, we find that the entanglement entropy no longer depends on the size of the subsystem. Instead, it converges to a constant value, indicating the transition into the so-called ``area-law" phase. In this phase, the entanglement entropy, asymptotically, scales as $\mathcal{O}(1)$. This scaling behavior is consistent with studies on monitored free fermion dynamics, which have also identified the absence of a volume-law phase for finite $p$ \cite{10.21468/SciPostPhys.7.2.024,Alberton_2021,PhysRevResearch.2.033017}. To distinguish this critical measurement rate from others explored later, we denote it as $p^{EE}_c$.

\begin{figure}
    \raggedright
    \includegraphics[width=0.45\textwidth]{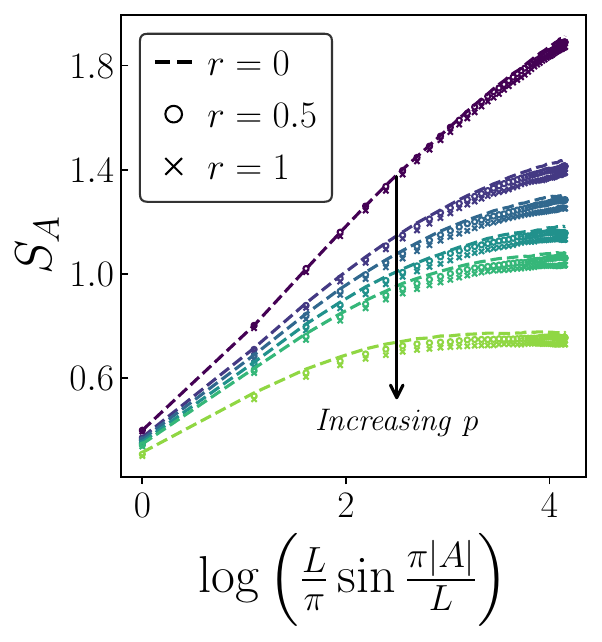}
    \caption{The entanglement entropy for contiguous subsystems, similar to \cref{fig:ents}, comparing the steady-state entanglement for $r=0$ (dashed lines), $r=0.5$ ($\circ$) and $r=1$ ($\cross$). The values of $p$ are 0.2, 0.23, 0.24, 0.25, 0.26 and 0.3. The scaling of $S_A$ with $|A|$ is insensitive to $r$, confirming that feedback has no effect on the measurement-induced entanglement phase transition.}
    \label{fig:ents_r}
    
\end{figure}

Next, we consider the scenario where $r<1$. Interestingly, we numerically find that the entanglement transition, characterized by $p_c^{EE}$, remains insensitive to the value of $r$ (\cref{fig:ents_r}). In other words, changing the feedback rate does not affect the critical measurement rate for the entanglement transition. Disregarding the classical flags for the moment, the application of SWAP gate between two sites, following projective measurements on those sites, does not change the entanglement of a wavefunction since, ultimately, the two sites undergoing a SWAP both host product states. It is less trivial to explain why the deactivation of sites via the classical flags appears to have no bearing on the MIPT. We return to this at the end of \cref{sec:C_Map}.

\subsection{Absorbing Transition}
\label{sec:abs_T}

In addition to the MIPT, we observe a transition in a more conventional observable -- the charge density-imbalance between neighboring sites $\Delta$ -- as $p$ is further increased. We first fix $r=1$. For small $p$,  $1 - \Delta(t)$ briefly decays before settling to its final value. However, as $p$ is increased beyond a critical rate $p^\Delta_c\sim 0.6$ (which is generally different from $p^{EE}_c$), $1 - \Delta$ decays diffusively as $1 - \Delta(t)\sim t^{-0.5}$, at which point, the system is said to be in an ``absorbing" phase. At $p=p^\Delta_c$, $1 - \Delta$ decays as $1 - \Delta(t)\sim t^{-0.286}$. This behavior is characteristic of the PC universality class of nonequilibrium dynamics, a well-studied classical stochastic process that involves the creation and annihilation of classical particles, but only in pairs, leading to a conservation of parity~\cite{hinrichsen2000non}.

\begin{figure*}
    \centering
    \includegraphics[width=0.45\textwidth]{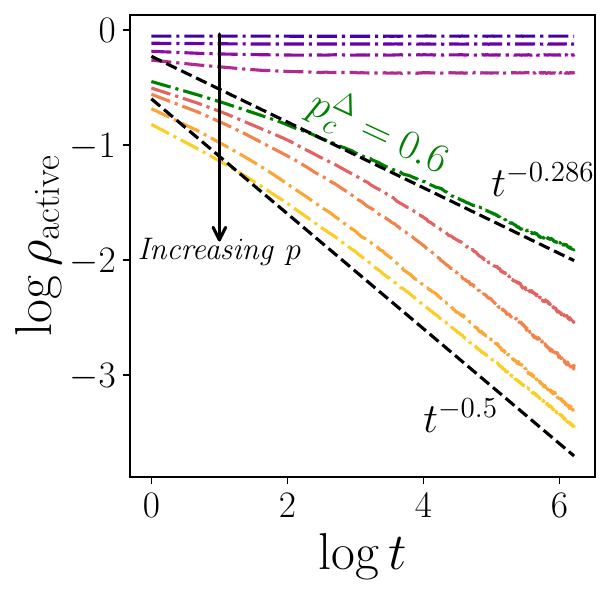}\hspace{0.05\textwidth}
    \includegraphics[width=0.45\textwidth]{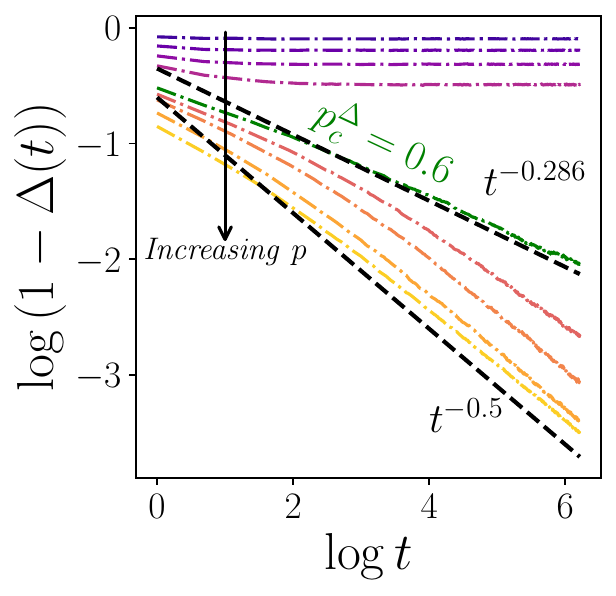}
    \caption{Time evolution of the density of active sites (above) and the charge imbalance $1-\Delta(t)$ (below), plotted on log-log axes, for a system with $L=200$ sites and $r=1$ held fixed. 
    For each plot, the nine curves correspond to $p=0.1$, 0.2, 0.3, 0.4, 0.6, 0.65, 0.7, 0.8, 0.9, arranged from top to bottom.
     An absorbing phase transition is observed as $p$ is varied, from an ``active" phase, where $\rho_{\rm active}$ saturates to a finite value, to an absorbing phase, where $\rho_{\rm active}$ decays diffusively to zero. These two regimes are separated by a power law decay of $\rho_{\rm active}(t)\sim t^{-0.286}$ at $p_c^{\Delta}=0.6$. The dynamics of $1-\Delta(t)$ and $\rho_{\rm active}$ are effectively identical, showing that $\rho_{\rm active}$ is a fitting diagnostic for the transition.}
    \label{fig:transition}
\end{figure*}

We also observe a nearly identical behavior in the dynamics of the density of active sites, as determined by the classical flag variables defined in \cref{sec:model}, shown in \cref{fig:transition}. When a state sufficiently close to the target state $\ket{101010\dots}$ is subject to the circuit, it will result in almost all flags being set to $0$, thereby minimizing the density of active sites, provided the measurement rate is high enough. Since $1 - \Delta$ is also minimized by this target state, it is expected that its behavior should be reflected in $\rho_{\rm active}$.

\begin{figure}
    \centering
    \includegraphics[width=0.475\textwidth]{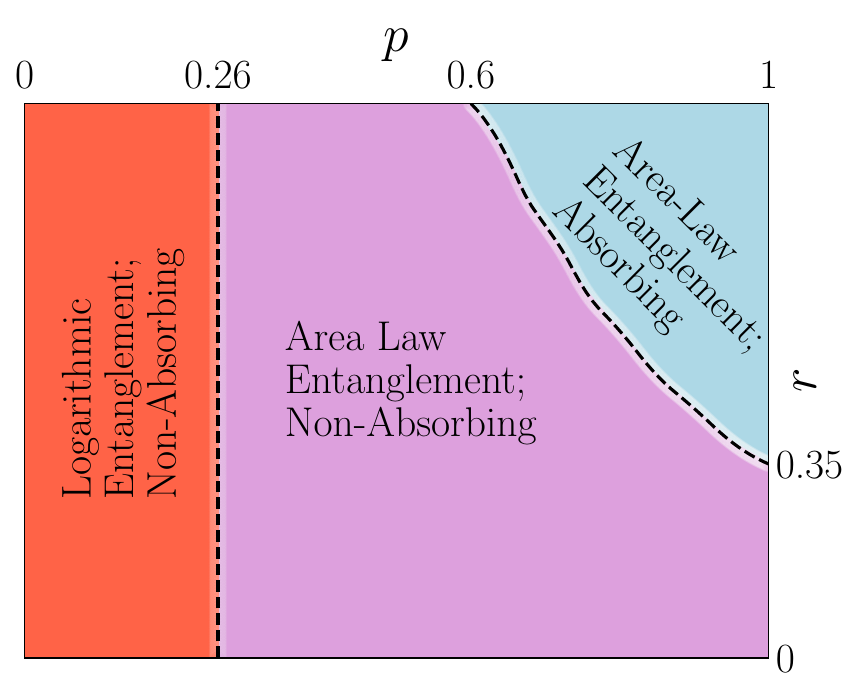}
    \caption{An estimation of the phase diagram for the model, showing the rough boundaries of both the absorbing and logarithmically entangled phase, as $p$ and $r$ are varied. The feedback rate $r$ has no effect on the scaling of the entanglement entropy, however the boundary between the absorbing and non-absorbing phases is a nonlinear curve in the $p-r$ space. In the absorbing phase, the final state is $\ket{101010\dots}$. }
    \label{fig:PhaseDiag}
\end{figure}

If instead we fix $p=1$ and vary $r$, we see a similar transition at $r^\Delta_c = 0.35$. In general, there exists a curve of values of $r$ and $p$ which separates the absorbing phase from the non-absorbing phase. This is to be expected, since (a) feedback can only be applied once a measurement is made, and (b) the steady state, averaged over measurement outcomes in the absence of feedback, is featureless and proportional to $\mathbb{I}$. The phase diagram for the model in $p-r$ space is shown in \cref{fig:PhaseDiag}.

The agreement between $1-\Delta$ and $\rho_{\rm active}$ is of special relevance to experimental realizations of this transition. Since measurements of the occupations of neighboring sites are performed as a part of the circuit, the outcomes of this measurements can be used to estimate the charge-imbalance between neighboring sites, reducing experimental overhead~\cite{ravindranath2022entanglement}. Since neighboring sites are deemed inactive only if they are in the $10$ or $01$ state, the flag density is itself an indicator of the success of our protocol in preparing the target state, and thus, an estimator for $1-\Delta$.

Two remarks on the existence of a steady-state are in order. In any generic system with a unique, absorbing phase with no special symmetries (excluding parity), all initial states will, independent of the parameter values $p$ and $r$, flow to this target state. The distinction between the two phases of an absorbing phase transition is crucially dependent on the timescales under consideration; in the active phase, the time taken to reach the absorbing state is $\mathcal{O}(e^L)$. Therefore, in order to distinguish the absorbing and non-absorbing phases, one must focus on time scales $t\sim\mathcal{O}(L^2)$ or more generally, at most polynomial in $L$. Since the phenomena that we are considering persist for thermodynamically large times, we refer to the behavior at $t\sim\bigO{L^2}$ as the ``steady state behavior". Therefore, when referring to specific observables (such as $S_A, \rho_{active}$ or $\Delta$), ``steady-state values" refers to their values at $t\sim\bigO{L^2}$, averaged over trajectories, while the ``steady-state" refers to the unique quantum state reached at $t\sim\bigO{L^2}$ (if it exists).

With these definitions, $\Delta$ reaches its steady-state value in $\bigO{1}$ time in the active phase. However, not every quantum trajectory converges to the same quantum state. In the absorbing phase, \textit{each} trajectory reaches the target state. Moreover, it is only in this phase that a steady state even exists at the timescales considered.

\subsection{Mapping to a Classical Model}
\label{sec:C_Map}

As shown in \cref{sec:abs_T}, the dynamical phase transition observed in
$\rho_{\rm active}$ and $\Delta(t)$ exhibits a striking resemblance to classical stochastic models within the PC non-equilibrium universality class~\cite{hinrichsen2000non}. In what follows, we aim to elucidate the underlying physics behind this phenomenon.

\begin{figure}[tbh]
    \centering
    \includegraphics[width=0.45\textwidth]{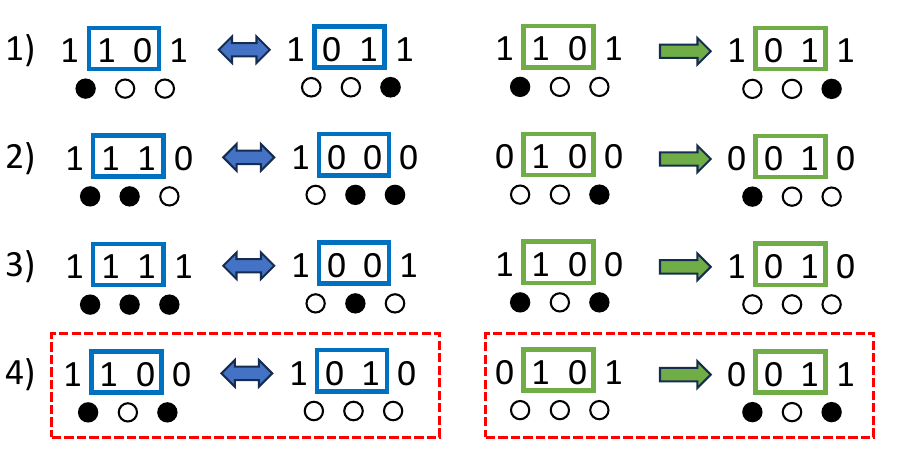}
    \caption{Update rules for the corresponding classical model, in the absence of classical flag variables $f_i$. These rules are shown for the even layers of the circuit. (Left) The blue boxes describe unitary gates. The update rules for the states not shown are straightforwardly obtained by flipping all 4 bits (e.g. the particle configuation $\bullet\circ\circ$ equally describes the bitstring 1101 or 0010). (Right) The green boxes denote the measurement and SWAP operations. The processes highlighted in red are those which map the vacuum state (the state with no $\bullet$ particles) to active states. States not shown here are not affected by the corrective operations. In the presence of flag variables, the feedback (green) process on sites $i$ and $i+1$ additionally labels them as inactive, setting $f_i = f_{i+1} = 0$.}
    \label{fig:cl_rules}
\end{figure}

We follow the method developed in Ref.~\cite{ravindranath2022entanglement}. For each individual quantum trajectory undergoing the non-unitary time evolution, at time $t$, its wave function can be written as a superposition of states in the occupation number basis 
\begin{align}
\ket{\psi (t)} = \sum\limits_{\qty{n_j}} c_{\qty{n_j}}(t) \ket{\qty{n_j}}.
\end{align}
Each basis state can be represented by a string of bits of length $L$ -- or bitstring -- representing its pattern of ferminonic occupations as $\ket{\qty{n_j}}$, with $n_1,n_2\dots n_L = 0, 1$. The superposition runs over all states of a given parity $\prod\limits_{i=1}^L (1-2\hat{n}_i) = \pm 1$ since the dynamics preserves the parity symmetry. Ignoring the amplitudes $c_{\qty{n_j}}$ for the moment, we seek to deduce which bitstrings can contribute to $\ket{\psi(t)}$ and how these bitstrings evolve with time. This can be explored by mapping the bitstring dynamics to a model of classical particles under stochastic dynamics.
The transition probability between bitstrings $\qty{n_j}$ and $\qty{n'_j}$ is determined by $|\mel{\qty{n_j}}{\widetilde{U}}{\qty{n'_j}}|^2\neq0$, for \textit{any} unit time-step realization of the circuit $\widetilde{U}$.

To understand the emergence of the PC universality class, we introduce particles defined on the bonds (i.e. between sites) of the lattice. Occupations of $(1,1)$ or $(0,0)$ across the $i^\text{th}$ bond are identified with a particle $\bullet$ at bond $i$, while $(1,0)$ or $(0,1)$ are identified with the absence of a particle or a hole, denoted as $\circ$. Particles and holes so defined can hop by two sites in either direction. A particle can also branch to create 2 additional particles to its left and right, and two particles can annihilate in pairs. The update rules are summarized in \cref{fig:cl_rules}, in the absence of flag variables $f_j$.

\begin{figure}
    \centering
    \includegraphics[width=0.45\textwidth]{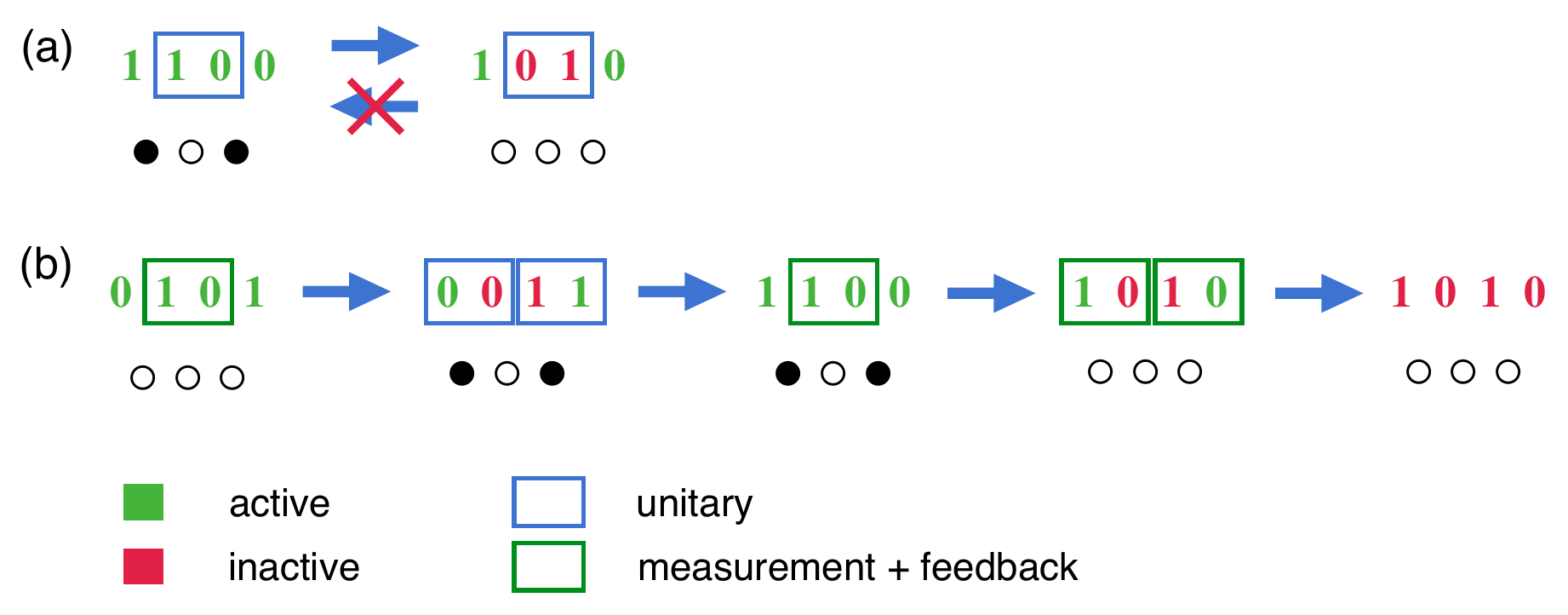}
    \caption{Combined dynamics of bitstring (fermion occupation) and classical flag. (a) The flag variable imposes a directionality on the unitary evolution, namely, unitary gates no longer act on two sites jointly once they are both labelled as inactive (colored in red). (b) Particle pairs created by measurement and feedback must eventually annihilate in pairs once the target state is reached. }
    \label{fig:flag}
\end{figure}

In this particle language, the desired target state $\ket{\psi_{\rm targ}} = \ket{1010\dots}$ is the empty state with no $\bullet$ particle. In order to successfully steer the initial configurations to this state, we require that $\ket{\psi_{\rm targ}}$ is an absorbing state, to which all other states eventually flow. However, the unitary process which creates a pair of particles (process 4 in \cref{fig:cl_rules}), being reversible by definition, cannot lead to an absorbing state, which requires a directionality to
time evolution. The introduction of the classical flags $f_j$ is specifically intended to fix this limitation.
According to the protocol described in Sec.~\ref{sec:model}, once a measurement is made and the local bitstring configuration agrees with that in $|\psi_{\rm targ}\rangle $, the flag variables are set to zero: $f_j = f_{j+1} = 0$, and unitary gates no longer act on these two sites jointly [Fig.~\ref{fig:flag}(a)].
Thus, the introduction of feedback and classical variables effectively impose a directionality on the dynamics that would be absent with unitary dynamics alone.

\begin{figure*}
    \centering
    \includegraphics[width=0.33\textwidth]{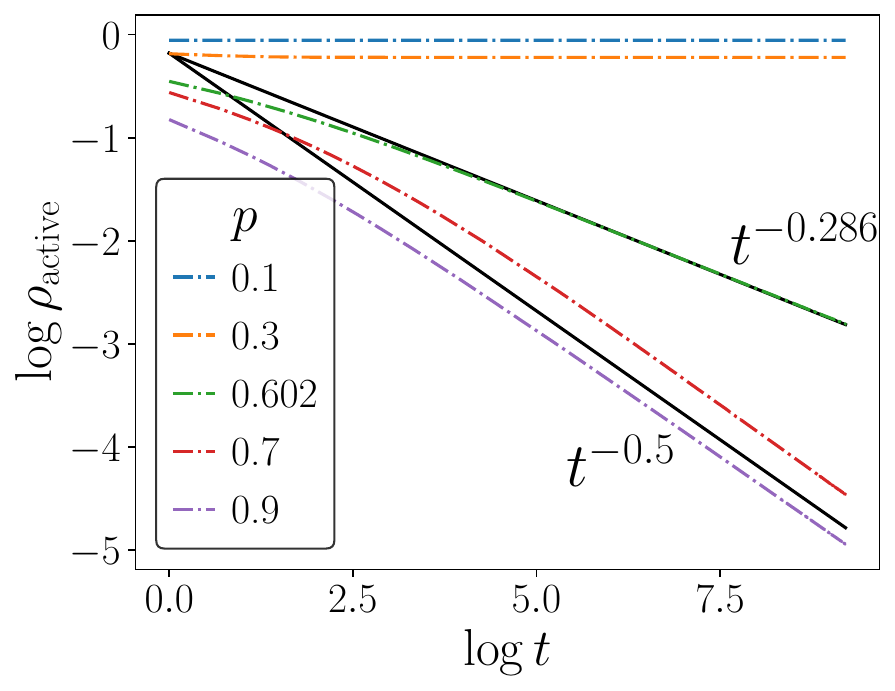}%
    \includegraphics[width=0.33\textwidth]{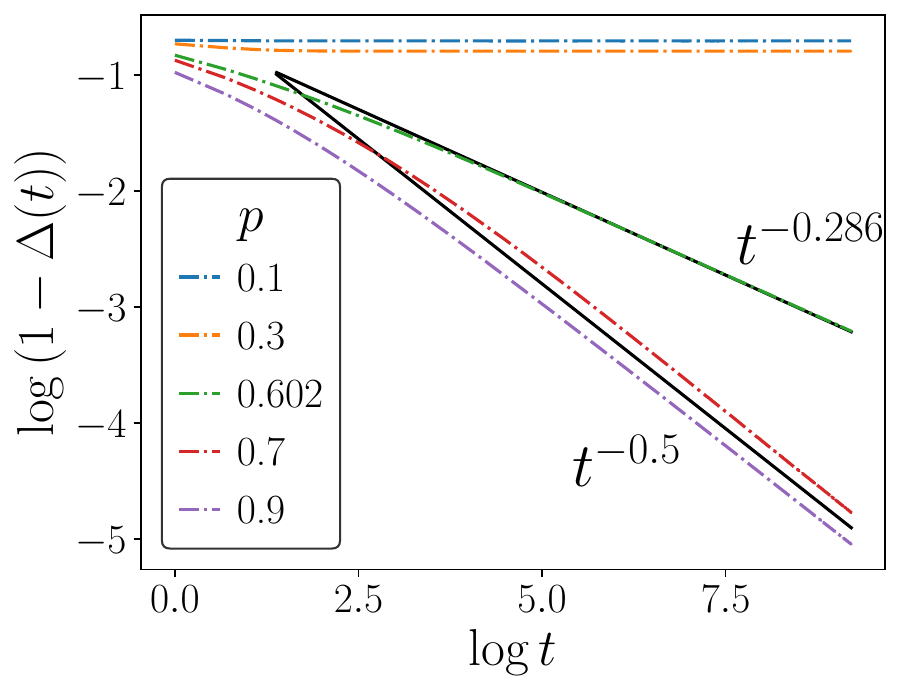}
    \includegraphics[width=0.33\textwidth]{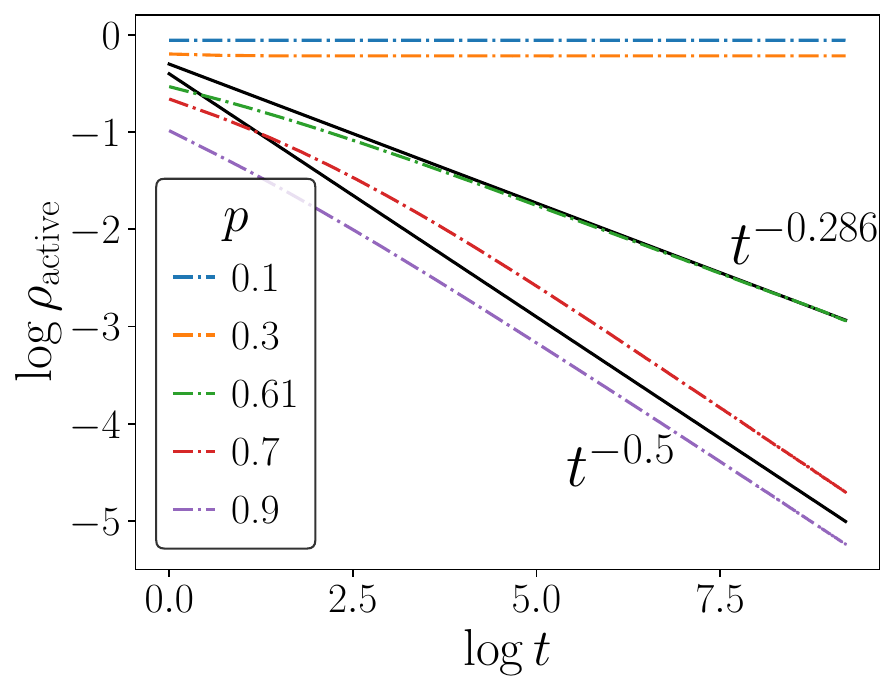}
    \caption{Simulation of the classical stochastic dynamics  following the same update rules as those listed in Fig.~\ref{fig:cl_rules}, with the addition of the flag variables which are updated analogously to the quantum model, on a lattice of $L=1000$ sites. The time evolution of the density of active sites (left) and the charge imbalance $\Delta$ (center) are in remarkable agreement with the full quantum simulation \cref{fig:transition}. The critical point $p_c^\Delta\sim 0.6$ is the same as in the quantum dynamics. (Right) Also shown is $\rho_{\text{active}}$ corresponding to the case where unitary gates and feedback act on 3 sites at a time, showing that neither $p_c$ nor the universality class change.}
    \label{fig:classical}
\end{figure*}

Feedback, on the other hand, also introduces a process which creates particles from the empty state (process 4 in \cref{fig:cl_rules}).  
With the introduction of classical variables, the target state is no longer characterized by the occupation pattern of the sites (or equivalently, the $\bullet$ particles) alone. Instead, one must also specify that in the target state, $f_i = 0$. Since sites are only labelled as inactive if they have the correct ordering of $(1,0)$, measurement process 4 eventually leads to process 3, where the resulting state has no particles, no active sites \textit{and} is invariant under feedback [Fig.~\ref{fig:flag}(b)]. 
Thus, feedback, in tandem with a classical labelling of sites, can create an absorbing state. Since measurements and feedback both drive the states towards $\ket{\psi_{\rm targ}}$, while the unitary gates randomize the occupation pattern, $\ket{\psi_{\rm targ}}$ eventually becomes an absorbing state at a sufficiently high rate of measurement and feedback.

To verify this picture, we numerically simulate this classical stochastic dynamics. In the classical simulation, the unitary gates and measurement processes are replaced by classical processes that follow the update rules in \cref{fig:cl_rules}, with the addition of the flag variables which are updated analogously to the quantum case. The dynamics of the density of active particle obtained from the classical simulation is shown in  \cref{fig:classical}, for various $p$. When $p<p_c$, the density of the particle saturates to a finite constant. On the other hand, when $p>p_c$, the particle density decays diffusively to zero. At the critical point, it decays as a power law function with an exponent consistent with PC universality class. We also observe a similar scaling behavior for the density of active sites in the flag variables. It is important to note that the parity symmetry is not obeyed by the number of active sites, yet their dynamics, influenced by the particle dynamics, still adhere to the PC universality class. Finally, we perform a detailed scaling analysis for various system sizes as well as for different $p$ in the vicinity of $p_c$ to extract the critical exponents, which are found to be in close agreement with those obtained from previous studies \cite{henkel2008non,hinrichsen2000non}. These results confirm that the combined dynamics of the classical stochastic process in Fig.~\ref{fig:cl_rules} and the classical flags has an absorbing phase transition belonging to the PC universality class.

\begin{figure}
    \hspace{-1em}%
    \includegraphics[width=0.25\textwidth]{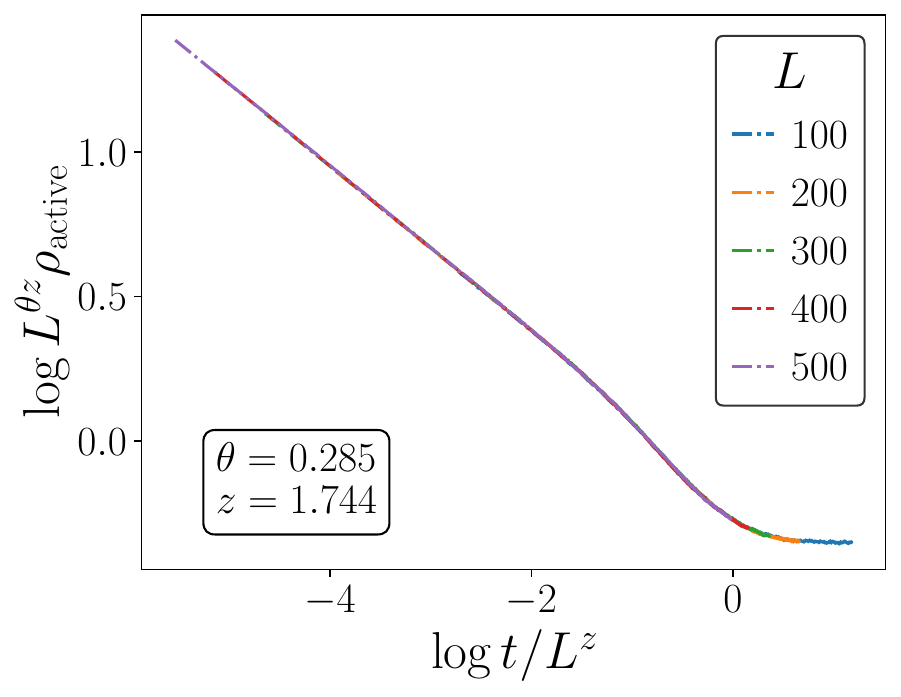}%
    \includegraphics[width=0.25\textwidth]{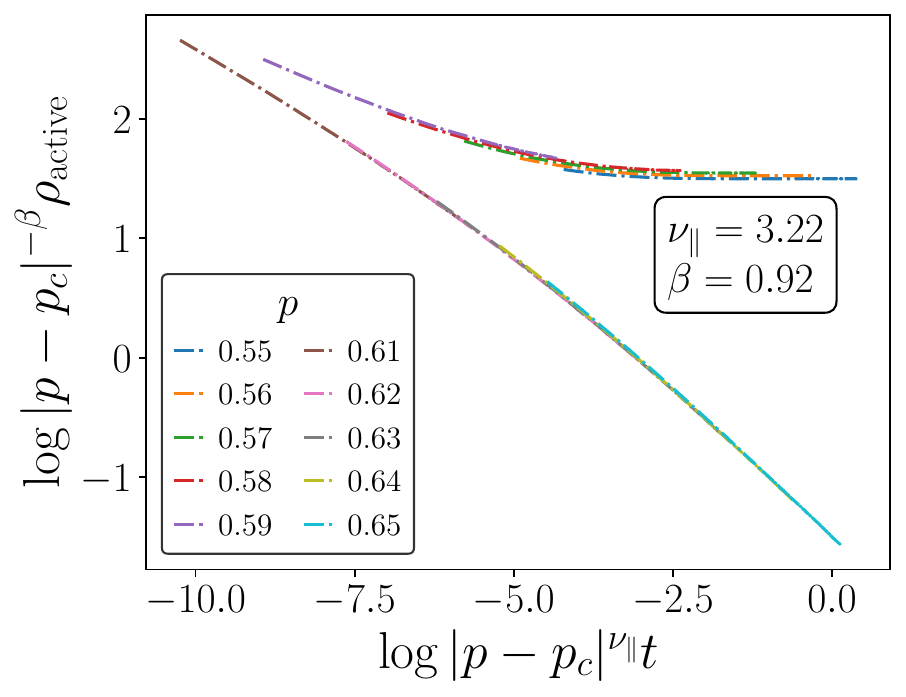}
    \caption{Scaling analyses, obtained for the classical simulation, for density of active sites $\rho_{\rm active}$ (left) at $p=p_c$ for various system sizes and (right) for $L=1000$ at different $p$ near $p_c$. In both cases, with $p_c=0.605$, excellent data collapses are observed, from which the critical exponents are extracted. In particular, we find $\beta = \nu_{\parallel} \theta$, consistent with the scaling hypothesis. Details of the scaling form are given in \cref{app:PC}.}
    \label{fig:scaling}
\end{figure}

At time $t$, the quantum wave function $\ket{\psi(t)}$ can only be spanned by the bitstrings that survive at time $t$. In the absorbing phase and at late times, the sole bitstring that survives at long times is $1010\dots10$, ensuring that the steady state is the product state $\ket{\psi_{\rm targ}}$. In the active phase, $\ket{\psi}$ instead is spanned by an extensive number of bitstrings. Thus, even though $\Delta$ reaches a steady-state value upon averaging, no quantum trajectory in the active phase settles to a specific steady state. Therefore the entanglement phase transition must occur within the active phase, as depicted in the phase diagram \cref{fig:phase} and confirmed numerically.

Armed with an understanding of the absorbing phase transition, we now address the counterintuitive observation that feedback does not affect the critical measurement rate $p_c$. We have already seen why the quantum operations involved in the feedback do not alter entanglement. The ineffectiveness of deactivation of sites at altering $p_c$ is more subtle. At the value of $p$ at which the MIPT is observed, $\rho_{\rm active}$ remains close to 1 even for $r=1$ (as in \cref{fig:classical}). It is also known (e.g.\ from classical simulations) that the dynamical processes described in \cref{sec:C_Map} cannot create large-scale fracturing of the system; configurations with inactive sites that break the system into O(1) pieces are highly improbable. Therefore, at these values of $p$ and $r$, we can assume that the entanglement dynamics will be the same as for a smaller system of size $L'\lesssim\rho_{active}L$. Thus, in the thermodynamic limit, the entanglement transition is not affected.

Lastly, the role of unitary gates in the classical mapping (\cref{fig:cl_rules}) is to enable \textit{reversible} transitions between bitstrings of the same parity. The phenomenology of the absorbing phase transition persists for any gate set that enables these transitions. Generic unitary gates, restricted only by parity symmetry, can do so.

\section{Discussions and outlook}

In this work, we have shown that an adaptive protocol can successfully prepare charge-density-wave states in free fermion systems. While no-go theorems exist which preclude the creation of ordered states in one dimension, our protocol is able to avoid this pitfall by introducing classical flags, doing away with the need for fine-tuned unitary gates~\cite{ravindranath2022entanglement,o2022entanglement}. Steering to this particular target state happens via a non-equilibrium phase transition which we find belonging to the PC universality class. 
In contrast to previous investigations~\cite{ravindranath2022entanglement,o2022entanglement,piroli2022triviality,PhysRevLett.130.120402}, inferring that the dynamical phase transition belongs to the PC universality class cannot be easily achieved solely based on a feedback mechanism or the presence of a classical flag. Instead, it arises intrinsically from the combined effect of fermionic parity in the quantum dynamics and the classical labelling.

 We have also conclusively demonstrated that the entanglement and absorbing phase transitions in generic, monitored free fermion systems are independent of each other.
In addition, the protocol described in the paper is not limited to the free fermion dynamics and can be easily generalized to fermionic systems with density-density interactions. When expressed in the occupation number basis, the fermion interaction term will introduce extra phases for each basis but will not affect the underlying classical bit string dynamics. As a consequence, the entanglement phase transition will be affected while the absorbing phase transition to ordered state remains the same.

A consideration borne out of practical importance is the robustness of this protocol to noise. When the projective measurements are replaced by the imperfect weak measurements, although the entanglement transition persists, the absorbing phase transition in general will be absent. We verify this numerically and observe that upon replacing projective measurements with weak measurements, the density of particles will always approach a finite constant at long times. We have also modelled depolarizing noise by considering random bit-flip errors in the classical dynamics. The transition is again washed out in this case. The reason for the absence of the transition in either case can be understood in the particle language: both these errors lead to the creation of $\bullet$ particles from the vacuum, thus destroying the absorbing nature of $\ket{\psi_{\rm targ}}$. In contrast, dephasing noise will not have an impact on this absorbing phase transition since it will not affect the underlying particle dynamics.

An interesting question for future work is the preparation of quantum states with different ordering patterns using fermionic adaptive quantum circuits.
For instance, we could prepare $|00\cdots 0\rangle$ by using a similar protocol introduced in this work. While fermionic parity appears to pose an unavoidable constraint on the preparation of certain target states, it would be worth exploring if the approach to these target states could be made faster than in a diffusive fashion. We would also like to extend this idea to prepare quantum states that exhibit more exotic entanglement patterns than simple product states.

\section*{Acknowledgments}

This research is partially supported by a start-up fund at Peking University (Z.-C.Y.) and Grant No.~12375027 from the National Natural Science Foundation of China (Z.-C.Y.). This research is also supported in part by the Google Research Scholar Program and is supported in part by the National Science Foundation under Grant No.~DMR-2219735 (V.R. and X.C.). We gratefully acknowledge computing resources from Research Services at Boston College and the assistance provided by Wei Qiu. 

\appendix

\section{Numerical Simulation of Gaussian States}
\label{app:numerics}

Free fermions or Fermionic Gaussian States (FGS) constitute an important class of quantum systems which can be efficiently simulated classically \cite{BravyiFLO,10.21468/SciPostPhysLectNotes.54}. The fundamental reason underlying this is the fact that FGS obey Wick's theorem, so any $n$-point correlation function can be completely determined from 2-point correlators of the form $\ev*{c^\dagger_ic^\dagger_j}$ and $\ev*{c^\dagger_i c_j}$. Any FGS $\ket{\psi}$ is then completely determined by the $C$ matrix, defined as $C =  
    \begin{pmatrix}
    \langle c_i c^\dagger_j \rangle & \expval{c_i c_j}  \\
    \langle c^\dagger_i c^\dagger_j \rangle & \langle c^\dagger_i c_j \rangle
    \end{pmatrix}.$ Thus, instead of tracking the evolution of a generic superposition in Fock space, which requires storing $\mathcal{O}\qty(2^L)$ variables, one can instead track the evolution of $C$ alone, which only requires $\mathcal{O}(L^2)$ variables to record, for a system of fermions on $L$ sites.

For numerical implementations, it is useful to note that an FGS $\ket{\psi}$ can always be completely determined by the $L$ fermionic operators -- labelled $d_1 \dots d_L$ -- which annihilate it~\cite{ravindranath2022robust,PhysRevResearch.2.033017}

\begin{equation}
    d_j\ket{\psi} = 0,\text{ for $j=1\dots L$}.
\end{equation}
Moreover, these operators can be expressed as linear combinations of the creation and annihilation operators, as 
\begin{equation}
    d_j = \sum\limits_{k=1}^L(\alpha)^*_{j,k} c_k + (\alpha)^*_{j,k+L} c^\dagger_k.
\end{equation} 
This defines a representation of each annihilation operator $d_j$ of $\ket{\psi}$ as a $2L\times1$ column vector having $\alpha_{j,k}$ as its $k^\text{th}$ entry. These vectors can be collected to form a $2L\times L$ matrix $\alpha$

\begin{equation}
    \alpha = \mqty(\vert&&\vert\\
    \alpha_1&\dots&\alpha_L\\
    \vert&&\vert)
\end{equation}

By defining the vector $\vec{c}$

\begin{equation}
\vec{c} = \mqty(c_1\\\vdots\\c_L\\c^\dagger_1\\\vdots\\c^\dagger_L),
\end{equation}
$d_j$ can be written as $d_j = \alpha_j^\dagger\vec{c}$. In this notation, the $ij$ element of the matrix is the $i^\text{th}$ entry of the column vector $\alpha_j$, so $\alpha_{i,j} \equiv \qty(\alpha_j)_i$.  The anticommutativity of the annihilation operators $\qty{d_j,d_k} = 0$ by virtue of being fermionic reflects in the orthonormality of the columns of $\alpha$, i.e. $\alpha_j^\dagger\alpha_k = \delta_{jk}$. The $C-$ matrix can be conveniently expressed as $C = \alpha\alpha^\dagger$, so it suffices to follow the evolution of $\alpha$ alone. The evolution of $\alpha$ under unitary operations and projective measurements is detailed below.

\textbf{Unitary Gates: } A fermionic gaussian unitary operation $U$ is 
one that can be written as an exponential of a hermitian operator that is quadratic in $\qty{c_j, c^\dagger_j}$. Such a $U$ has the form
\begin{equation}
    U = \exp\qty(-i\frac{\vec{c}^\dagger \mathcal{H}\vec{c}}{2})
\end{equation}
with $\mathcal{H}^\dagger = \mathcal{H}$. $\mathcal{H}$ actually has more structure than a generic Hermitian matrix, and can without loss of generality be written in terms of two $L\times L$ block matrices $A$ and $B$
\begin{equation*}
    \mathcal{H} = \mqty(A&B\\
    B^\dagger&-A^T),
\end{equation*}
where $A^\dagger = A$ and $B^T = -B$.

If $\ket{\psi}\to U\ket{\psi}$, \[\alpha\to \exp(-i\mathcal{H}) \alpha\].

\textbf{Projective Measurements: }For concreteness, we focus on the measurement of the occupation of site $i$ $n_i = c^\dagger_i c_i$. Other Gaussian measurements can be straightforwadly implemented by applying a unitary rotation to the state prior to and after a measurement of the occupation number. When $n_i$ of a state $\ket{\psi}$ is measured, the post-measurement state $\ket{\psi'}$ is

\begin{equation}
    \ket{\psi}\to\ket{\psi'} = \begin{cases}
        \frac{n_i\ket{\psi}}{\sqrt{\ev{n_i}{\psi}}},&\text{ with probability} \ev{n_i}{\psi}\\
        \frac{(1- n_i)\ket{\psi}}{\sqrt{\ev{1 - n_i}{\psi}}},&\text{ with probability} \ev{1 - n_i}{\psi}\\
    \end{cases}.
\end{equation}
In the first case, the state is guaranteed to have a particle on site $i$, that is, $\ev{n_i}=1$, after measurement. We explicitly detail the implementation of such a measurement. Given $d_1\dots d_L$ which annihilate $\ket{\psi}$, the objective is to find the operators $\qty{\widetilde{d}_j\equiv \widetilde{\alpha}_j^\dagger \vec{c}}$ which annihilate $\ket{\psi'}\propto n_i\ket{\psi}$. It is instructive to first determine the action of any $d_j$ on $\ket{\psi'}$. For future convenience, we break $d_j$ into two parts $d^{\|}_j = (\alpha_j)^*_i c_i + (\alpha_j)^*_{i+L} c^\dagger_i$ and $d^\perp_j = d_j - d^{\|}_j$, noting that 

\begin{equation}
    \begin{aligned}
     \comm{d^\perp_j}{n_i} &= 0\\
     d^{\|}_j n_i &= (\alpha_j)^*_i c_i\\
     n_i d^{\|}_j &= (\alpha_j)^*_{i+L} c^\dagger_i\\
    \end{aligned}
\end{equation}

Since $d_j\ket{\psi} = 0 \implies d^\perp_j\ket{\psi} = -d^{\|}_j\ket{\psi}$,

\begin{equation}
    \begin{aligned}
        d_j n_i \ket{\psi} = n_i d^\perp_j\ket{\psi} + d^{\|}_jn_i\ket{\psi}\\
        = (-n_i d^{\|}_j + d^{\|}_j n_i )\ket{\psi}\\
        = \qty[(\alpha_j)^*_i c_i - (\alpha_j)^*_{i+L} c^\dagger_i] \ket{\psi}.
    \end{aligned}
    \label{eq:ann_FGS}
\end{equation}

We now use the fact that any linear combination of annihilation operators of a state also annihilates that state to create (non-canonical) annihilation operators $d'_j \equiv \alpha'^\dagger_j \vec{c} = d_j - \frac{\qty(\alpha_{j})_{i+L}}{\qty(\alpha_{i_0})_{i+L}} d_{i_0}$. Since $\ev{n_i}{\psi}\neq 0$ by assumption, $\ev{n_i}{\psi}\neq 0\implies C_{i+L,i+L}\neq 0 \implies \sum_k\alpha_{i+L,k}\alpha^*_{i+L,k} = \sum_k|\alpha_{i+L,k}|^2\neq 0$ ensures the existence of at least one such $i_0$ with $\qty(\alpha_{i_0})_{i+L}\neq0$. Moreover, $\qty(\alpha'_j)_{i+L} = 0$, by construction. Applying \cref{eq:ann_FGS} to the operator $d'_j$ for $j\neq i_0$, we have

\begin{equation}
    d'_j n_i \ket{\psi} = (\alpha'_j)^*_i c_i \ket{\psi}
\end{equation}

An effect of measuring $n_i$ and finding the outcome $n_i = 1$ is that none of the resulting annihilation operators can be supported on $c_i$, as $\ev{n_i}{\psi}=1\implies \ev{c_ic^\dagger_i}{\psi} = 0 \implies C_{i,i} = 0 \implies \sum_k\alpha_{i,k}\alpha^*_{i,k} = \sum_k|\alpha_{i,k}|^2 = 0$. Thus, for all $j\neq0$, $(\alpha'_j)_i$ is set to 0. In doing so, we obtain $L-1$ annihilation operators for the post-measurement state. The last annihilation operator can be obtained by noting that if $\ev{n_i}{\psi'} = 1$, then $c^\dagger_i\ket{\psi} = 0$ by the Pauli exclusion principle. Thus, in addition to the $L-1$ operators, $c^\dagger_i$ also annihilates the state $\ket{\psi'}$.

Summarizing this calculation, the algorithm to obtain $\tilde{\alpha}$ following a measurement of $n_i$ which yields an outcome $1$ is:

\begin{enumerate}
    \item Define $i_0$ such that $|\qty(\alpha_{i_0})_{i+L}| \neq 0 \geq |\qty(\alpha_{j})_{i+L}|$ for $j=1\dots L$.
    \item For $j\neq i_0$, $\alpha_j \to \alpha'_j = \alpha_j - \frac{\qty(\alpha_{j})_{i+L}}{\qty(\alpha_{i_0})_{i+L}}\alpha_{i_0}$.
    \item For $j\neq i_0$, $\qty(\alpha'_j)_i\coloneqq0$
    \item $\qty(\alpha'_{i_0})_j\coloneqq \delta_{j,i+L}$
\end{enumerate}
$\widetilde{\alpha}$ is finally obtained by orthonormalizing (e.g. using a Gram-Schmidt process) the $L$ vectors $\alpha'$ and collecting them in a $2L\times L$ matrix. For completeness, if instead the post-measurement state has $\ev{n_i} = 0$, the algorithm can be obtained by enacting the modification $i\leftrightarrow i+L$. An alternate derivation of this can be obtained by considering weak measurements with operators of the form $e^{\pm\beta n_i}$ in the limit $\beta\to\infty$. The implementation of weak measurements is discussed in detail in \cite{ravindranath2022robust}.
\section{Overview of the PC Universality Class}
\label{app:PC}

A paradigmatic example of the parity conserving universality class is given by the branching-annihilating random walk (BARW), which consists of particles performing an unbiased random walk on a one-dimensional lattice of $L$ sites, where each site can be occupied by at most one particle. When all particles belong to a single species $A$, the BARW is defined by the following update rules, applied at each time step, to each particle (in addition to the rules defined by the random walk). A particle can ``branch" by giving rise to two offspring, and pairs of particles can annihilate each other upon contact, at a rate $q$:
\begin{equation}
    \begin{aligned}
        A &\to 3A\\
        A + A &\to \emptyset \quad \text{with probability} \ q.
    \end{aligned}
    \label{eq:BARWrule}
\end{equation}

The particle number changes with each update only by multiples of 2, hence conserving the parity of the particle number (even or odd). A dynamical phase transition is observed in the particle density $n(t) = \frac{1}{L}\sum\limits_{i=1}^L n_i(t)$ -- where $n_i(t)$ denotes the occupation of site $i$ at time $t$ -- as $q$ is varied. Beginning from an initial state with a finite density of particles, for $q<q_c$, $n(t)$ briefly decays, before saturating to a non-zero $q$-dependent constant. When $q$ is increased above $q_c$, the density decays diffusively as $n(t)\sim t^{-0.5}$. The absence of exponential decay is explained by the constraints imposed by parity conservation, since particles can only be annihilated in pairs. At the critical point $q_c$, the particle density also decays as a power law, but subdiffusively -- $n(t)\sim t^{-0.286}$. In the vicinity of the critical point, the density obeys the one parameter scaling form \cite{henkel2008non,hinrichsen2000non}

\begin{equation}
    n(t, L, |q-q_c|) = \lambda^{-\beta} f\qty(\lambda^{-\nu_{\|}}t, \lambda^{-\nu_{\perp}}L, \lambda|q-q_c|),
    \label{eq:fss1}
\end{equation}
for any real $\lambda$, at time $t$ for a lattice of size $L$. $\nu_{\|}$ and $\nu_{\perp}$ are the exponents which determine the scaling of the temporal and spatial correlation lengths respectively. Their ratio $z = \nu_{\|}/\nu_{\perp}$ is the dynamical exponent of the theory, known to be $1.744$ at $q=q_c$. It is useful to define another exponent $\theta = \beta/\nu_{\|}$, since at $q=q_c, n(t)\sim t^{-\theta}$. Additionally, the power-law decay of the density above $q_c$ dictates that a finite-size scaling form can be obtained at fixed $q>q_c$, but with $\theta=0.5$ and $z=2$, with this dynamical exponent $z$ being characteristic of diffusive dynamics. In the main text, we adopt the above scaling hypothesis Eq.~(\ref{eq:fss1}) where we identify $\rho_{\rm active}(t)$ as $n(t)$. In particular, we consider (1) the scaling of $\rho_{\rm active}(t)$ with $t$ and $L$ at the critical point (by choosing $\lambda=L^{1/\nu_\perp}$) and (2) the scaling of $\rho_{\rm active}(t)$ with $p$ and $t$ in the vicinity of $p_c$ and for $L\gg t^{1/z}$ (by setting $\lambda=1/|q-q_c|$). The scaling form~(\ref{eq:fss1}) implies that in the first case, $L^{z\theta}\rho_{\rm active}$ is a universal function of $\frac{t}{L^z}$, and in the second case, $\rho_{\rm active} |p-p_c|^{-\nu_{\parallel}\theta} \equiv \rho_{\rm active} |p-p_c|^{-\beta}$ is a universal function of $|p-p_c|t^{1/\nu_{\parallel}}$.

\bibliographystyle{quantum}
\bibliography{biblio_quantum}

\begin{thebibliography}{10}

\bibitem{fisher2023random}
Matthew~P.A. Fisher, Vedika Khemani, Adam Nahum, and Sagar Vijay.
\newblock ``Random quantum circuits''.
\newblock
  \href{https://dx.doi.org/10.1146/annurev-conmatphys-031720-030658}{Annual
  Review of Condensed Matter Physics {\bf 14}, 335--379}~(2023).

\bibitem{PhysRevLett.123.210603}
Aaron~J. Friedman, Amos Chan, Andrea De~Luca, and J.~T. Chalker.
\newblock ``Spectral statistics and many-body quantum chaos with conserved
  charge''.
\newblock \href{https://dx.doi.org/10.1103/PhysRevLett.123.210603}{Phys. Rev.
  Lett. {\bf 123}, 210603}~(2019).

\bibitem{PhysRevX.8.041019}
Amos Chan, Andrea De~Luca, and J.~T. Chalker.
\newblock ``Solution of a minimal model for many-body quantum chaos''.
\newblock \href{https://dx.doi.org/10.1103/PhysRevX.8.041019}{Phys. Rev. X {\bf
  8}, 041019}~(2018).

\bibitem{PhysRevLett.121.060601}
Amos Chan, Andrea De~Luca, and J.~T. Chalker.
\newblock ``Spectral statistics in spatially extended chaotic quantum many-body
  systems''.
\newblock \href{https://dx.doi.org/10.1103/PhysRevLett.121.060601}{Phys. Rev.
  Lett. {\bf 121}, 060601}~(2018).

\bibitem{PhysRevX.8.021014}
Adam Nahum, Sagar Vijay, and Jeongwan Haah.
\newblock ``Operator spreading in random unitary circuits''.
\newblock \href{https://dx.doi.org/10.1103/PhysRevX.8.021014}{Phys. Rev. X {\bf
  8}, 021014}~(2018).

\bibitem{PhysRevX.8.021013}
C.~W. von Keyserlingk, Tibor Rakovszky, Frank Pollmann, and S.~L. Sondhi.
\newblock ``Operator hydrodynamics, otocs, and entanglement growth in systems
  without conservation laws''.
\newblock \href{https://dx.doi.org/10.1103/PhysRevX.8.021013}{Phys. Rev. X {\bf
  8}, 021013}~(2018).

\bibitem{PhysRevResearch.2.033032}
Pieter~W. Claeys and Austen Lamacraft.
\newblock ``Maximum velocity quantum circuits''.
\newblock \href{https://dx.doi.org/10.1103/PhysRevResearch.2.033032}{Phys. Rev.
  Res. {\bf 2}, 033032}~(2020).

\bibitem{PhysRevX.8.031057}
Vedika Khemani, Ashvin Vishwanath, and David~A. Huse.
\newblock ``Operator spreading and the emergence of dissipative hydrodynamics
  under unitary evolution with conservation laws''.
\newblock \href{https://dx.doi.org/10.1103/PhysRevX.8.031057}{Phys. Rev. X {\bf
  8}, 031057}~(2018).

\bibitem{PhysRevX.7.031016}
Adam Nahum, Jonathan Ruhman, Sagar Vijay, and Jeongwan Haah.
\newblock ``Quantum entanglement growth under random unitary dynamics''.
\newblock \href{https://dx.doi.org/10.1103/PhysRevX.7.031016}{Phys. Rev. X {\bf
  7}, 031016}~(2017).

\bibitem{PhysRevB.99.174205}
Tianci Zhou and Adam Nahum.
\newblock ``Emergent statistical mechanics of entanglement in random unitary
  circuits''.
\newblock \href{https://dx.doi.org/10.1103/PhysRevB.99.174205}{Phys. Rev. B
  {\bf 99}, 174205}~(2019).

\bibitem{PhysRevX.10.031066}
Tianci Zhou and Adam Nahum.
\newblock ``Entanglement membrane in chaotic many-body systems''.
\newblock \href{https://dx.doi.org/10.1103/PhysRevX.10.031066}{Phys. Rev. X
  {\bf 10}, 031066}~(2020).

\bibitem{mi2021information}
Xiao Mi, Pedram Roushan, Chris Quintana, Salvatore Mandrà, Jeffrey Marshall,
  Charles Neill, Frank Arute, Kunal Arya, Juan Atalaya, Ryan Babbush, Joseph~C.
  Bardin, Rami Barends, Joao Basso, Andreas Bengtsson, Sergio Boixo, Alexandre
  Bourassa, Michael Broughton, Bob~B. Buckley, David~A. Buell, Brian Burkett,
  Nicholas Bushnell, Zijun Chen, Benjamin Chiaro, Roberto Collins, William
  Courtney, Sean Demura, Alan~R. Derk, Andrew Dunsworth, Daniel Eppens,
  Catherine Erickson, Edward Farhi, Austin~G. Fowler, Brooks Foxen, Craig
  Gidney, Marissa Giustina, Jonathan~A. Gross, Matthew~P. Harrigan, Sean~D.
  Harrington, Jeremy Hilton, Alan Ho, Sabrina Hong, Trent Huang, William~J.
  Huggins, L.~B. Ioffe, Sergei~V. Isakov, Evan Jeffrey, Zhang Jiang, Cody
  Jones, Dvir Kafri, Julian Kelly, Seon Kim, Alexei Kitaev, Paul~V. Klimov,
  Alexander~N. Korotkov, Fedor Kostritsa, David Landhuis, Pavel Laptev, Erik
  Lucero, Orion Martin, Jarrod~R. McClean, Trevor McCourt, Matt McEwen, Anthony
  Megrant, Kevin~C. Miao, Masoud Mohseni, Shirin Montazeri, Wojciech
  Mruczkiewicz, Josh Mutus, Ofer Naaman, Matthew Neeley, Michael Newman,
  Murphy~Yuezhen Niu, Thomas~E. O’Brien, Alex Opremcak, Eric Ostby, Balint
  Pato, Andre Petukhov, Nicholas Redd, Nicholas~C. Rubin, Daniel Sank, Kevin~J.
  Satzinger, Vladimir Shvarts, Doug Strain, Marco Szalay, Matthew~D.
  Trevithick, Benjamin Villalonga, Theodore White, Z.~Jamie Yao, Ping Yeh, Adam
  Zalcman, Hartmut Neven, Igor Aleiner, Kostyantyn Kechedzhi, Vadim
  Smelyanskiy, and Yu~Chen.
\newblock ``Information scrambling in quantum circuits''.
\newblock \href{https://dx.doi.org/10.1126/science.abg5029}{Science {\bf 374},
  1479--1483}~(2021).

\bibitem{PhysRevLett.127.230602}
Hansveer Singh, Brayden~A. Ware, Romain Vasseur, and Aaron~J. Friedman.
\newblock ``Subdiffusion and many-body quantum chaos with kinetic
  constraints''.
\newblock \href{https://dx.doi.org/10.1103/PhysRevLett.127.230602}{Phys. Rev.
  Lett. {\bf 127}, 230602}~(2021).

\bibitem{PhysRevB.100.134306}
Yaodong Li, Xiao Chen, and Matthew P.~A. Fisher.
\newblock ``Measurement-driven entanglement transition in hybrid quantum
  circuits''.
\newblock \href{https://dx.doi.org/10.1103/PhysRevB.100.134306}{Phys. Rev. B
  {\bf 100}, 134306}~(2019).

\bibitem{PhysRevB.98.205136}
Yaodong Li, Xiao Chen, and Matthew P.~A. Fisher.
\newblock ``Quantum zeno effect and the many-body entanglement transition''.
\newblock \href{https://dx.doi.org/10.1103/PhysRevB.98.205136}{Phys. Rev. B
  {\bf 98}, 205136}~(2018).

\bibitem{PhysRevB.99.224307}
Amos Chan, Rahul~M. Nandkishore, Michael Pretko, and Graeme Smith.
\newblock ``Unitary-projective entanglement dynamics''.
\newblock \href{https://dx.doi.org/10.1103/PhysRevB.99.224307}{Phys. Rev. B
  {\bf 99}, 224307}~(2019).

\bibitem{PhysRevX.9.031009}
Brian Skinner, Jonathan Ruhman, and Adam Nahum.
\newblock ``Measurement-induced phase transitions in the dynamics of
  entanglement''.
\newblock \href{https://dx.doi.org/10.1103/PhysRevX.9.031009}{Phys. Rev. X {\bf
  9}, 031009}~(2019).

\bibitem{noel2022measurement}
Crystal Noel, Pradeep Niroula, Daiwei Zhu, Andrew Risinger, Laird Egan,
  Debopriyo Biswas, Marko Cetina, Alexey~V Gorshkov, Michael~J Gullans, David~A
  Huse, et~al.
\newblock ``Measurement-induced quantum phases realized in a trapped-ion
  quantum computer''.
\newblock \href{https://dx.doi.org/10.1038/s41567-022-01619-7}{Nature Physics
  {\bf 18}, 760--764}~(2022).

\bibitem{koh2022experimental}
Jin~Ming Koh, Shi-Ning Sun, Mario Motta, and Austin~J. Minnich.
\newblock ``Measurement-induced entanglement phase transition on a
  superconducting quantum processor with mid-circuit readout''.
\newblock \href{https://dx.doi.org/10.1038/s41567-023-02076-6}{Nature Physics
  {\bf 19}, 1314--1319}~(2023).

\bibitem{hoke2023quantum}
Google~Quantum AI and Collaborators.
\newblock ``Measurement-induced entanglement and teleportation on a noisy
  quantum processor''.
\newblock \href{https://dx.doi.org/10.1038/s41586-023-06505-7}{Nature {\bf
  622}, 481--486}~(2023).

\bibitem{PhysRevLett.86.5188}
Robert Raussendorf and Hans~J. Briegel.
\newblock ``A one-way quantum computer''.
\newblock \href{https://dx.doi.org/10.1103/PhysRevLett.86.5188}{Phys. Rev.
  Lett. {\bf 86}, 5188--5191}~(2001).

\bibitem{PhysRevA.68.022312}
Robert Raussendorf, Daniel~E. Browne, and Hans~J. Briegel.
\newblock ``Measurement-based quantum computation on cluster states''.
\newblock \href{https://dx.doi.org/10.1103/PhysRevA.68.022312}{Phys. Rev. A
  {\bf 68}, 022312}~(2003).

\bibitem{tantivasadakarn2021long}
Nathanan Tantivasadakarn, Ryan Thorngren, Ashvin Vishwanath, and Ruben
  Verresen.
\newblock ``Long-range entanglement from measuring symmetry-protected
  topological phases''.
\newblock \href{https://dx.doi.org/10.1103/PhysRevX.14.021040}{Phys. Rev. X
  {\bf 14}, 021040}~(2024).

\bibitem{PRXQuantum.3.040337}
Tsung-Cheng Lu, Leonardo~A. Lessa, Isaac~H. Kim, and Timothy~H. Hsieh.
\newblock ``Measurement as a shortcut to long-range entangled quantum matter''.
\newblock \href{https://dx.doi.org/10.1103/PRXQuantum.3.040337}{PRX Quantum
  {\bf 3}, 040337}~(2022).

\bibitem{iqbal2023topological}
Mohsin Iqbal, Nathanan Tantivasadakarn, Thomas~M. Gatterman, Justin~A. Gerber,
  Kevin Gilmore, Dan Gresh, Aaron Hankin, Nathan Hewitt, Chandler~V. Horst,
  Mitchell Matheny, Tanner Mengle, Brian Neyenhuis, Ashvin Vishwanath, Michael
  Foss-Feig, Ruben Verresen, and Henrik Dreyer.
\newblock ``Topological order from measurements and feed-forward on a trapped
  ion quantum computer''.
\newblock \href{https://dx.doi.org/10.1038/s42005-024-01698-3}{Communications
  Physics {\bf 7}, 205}~(2024).

\bibitem{hauser2023continuous}
Jacob Hauser, Yaodong Li, Sagar Vijay, and Matthew P.~A. Fisher.
\newblock ``Continuous symmetry breaking in adaptive quantum dynamics''.
\newblock \href{https://dx.doi.org/10.1103/PhysRevB.109.214305}{Phys. Rev. B
  {\bf 109}, 214305}~(2024).

\bibitem{buchhold2022revealing}
M.~Buchhold, T.~Müller, and S.~Diehl.
\newblock ``Revealing measurement-induced phase transitions by
  pre-selection''~(2022).
\newblock  \href{http://arxiv.org/abs/2208.10506}{arXiv:2208.10506}.

\bibitem{iadecola2022dynamical}
Thomas Iadecola, Sriram Ganeshan, J.~H. Pixley, and Justin~H. Wilson.
\newblock ``Measurement and feedback driven entanglement transition in the
  probabilistic control of chaos''.
\newblock \href{https://dx.doi.org/10.1103/PhysRevLett.131.060403}{Phys. Rev.
  Lett. {\bf 131}, 060403}~(2023).

\bibitem{ravindranath2022entanglement}
Vikram Ravindranath, Yiqiu Han, Zhi-Cheng Yang, and Xiao Chen.
\newblock ``Entanglement steering in adaptive circuits with feedback''.
\newblock \href{https://dx.doi.org/10.1103/PhysRevB.108.L041103}{Phys. Rev. B
  {\bf 108}, L041103}~(2023).

\bibitem{o2022entanglement}
Nicholas O'Dea, Alan Morningstar, Sarang Gopalakrishnan, and Vedika Khemani.
\newblock ``Entanglement and absorbing-state transitions in interactive quantum
  dynamics''.
\newblock \href{https://dx.doi.org/10.1103/PhysRevB.109.L020304}{Phys. Rev. B
  {\bf 109}, L020304}~(2024).

\bibitem{PhysRevLett.130.120402}
Piotr Sierant and Xhek Turkeshi.
\newblock ``Controlling entanglement at absorbing state phase transitions in
  random circuits''.
\newblock \href{https://dx.doi.org/10.1103/PhysRevLett.130.120402}{Phys. Rev.
  Lett. {\bf 130}, 120402}~(2023).

\bibitem{piroli2022triviality}
Lorenzo Piroli, Yaodong Li, Romain Vasseur, and Adam Nahum.
\newblock ``Triviality of quantum trajectories close to a directed percolation
  transition''.
\newblock \href{https://dx.doi.org/10.1103/PhysRevB.107.224303}{Phys. Rev. B
  {\bf 107}, 224303}~(2023).

\bibitem{friedman2022measurement}
Aaron~J. Friedman, Oliver Hart, and Rahul Nandkishore.
\newblock ``Measurement-induced phases of matter require feedback''.
\newblock \href{https://dx.doi.org/10.1103/PRXQuantum.4.040309}{PRX Quantum
  {\bf 4}, 040309}~(2023).

\bibitem{Alberton_2021}
O.~Alberton, M.~Buchhold, and S.~Diehl.
\newblock ``Entanglement transition in a monitored free-fermion chain: From
  extended criticality to area law''.
\newblock \href{https://dx.doi.org/10.1103/PhysRevLett.126.170602}{Phys. Rev.
  Lett. {\bf 126}, 170602}~(2021).

\bibitem{PhysRevX.11.041004}
M.~Buchhold, Y.~Minoguchi, A.~Altland, and S.~Diehl.
\newblock ``Effective theory for the measurement-induced phase transition of
  dirac fermions''.
\newblock \href{https://dx.doi.org/10.1103/PhysRevX.11.041004}{Phys. Rev. X
  {\bf 11}, 041004}~(2021).

\bibitem{Turkeshi_2021}
Xhek Turkeshi, Alberto Biella, Rosario Fazio, Marcello Dalmonte, and Marco
  Schir\'o.
\newblock ``Measurement-induced entanglement transitions in the quantum ising
  chain: From infinite to zero clicks''.
\newblock \href{https://dx.doi.org/10.1103/PhysRevB.103.224210}{Phys. Rev. B
  {\bf 103}, 224210}~(2021).

\bibitem{Turkeshi_2022}
Xhek Turkeshi, Marcello Dalmonte, Rosario Fazio, and Marco Schir\`o.
\newblock ``Entanglement transitions from stochastic resetting of non-hermitian
  quasiparticles''.
\newblock \href{https://dx.doi.org/10.1103/PhysRevB.105.L241114}{Phys. Rev. B
  {\bf 105}, L241114}~(2022).

\bibitem{hinrichsen2000non}
Haye Hinrichsen.
\newblock ``Non-equilibrium critical phenomena and phase transitions into
  absorbing states''.
\newblock \href{https://dx.doi.org/10.1080/00018730050198152}{Advances in
  Physics {\bf 49}, 815--958}~(2000).

\bibitem{10.21468/SciPostPhys.7.2.024}
Xiangyu Cao, Antoine Tilloy, and Andrea~De Luca.
\newblock ``{Entanglement in a fermion chain under continuous monitoring}''.
\newblock \href{https://dx.doi.org/10.21468/SciPostPhys.7.2.024}{SciPost Phys.
  {\bf 7}, 024}~(2019).

\bibitem{BravyiFLO}
Sergey Bravyi.
\newblock ``Lagrangian representation for fermionic linear optics''.
\newblock \href{https://dx.doi.org/10.5555/2011637.2011640}{Quantum Info.
  Comput. {\bf 5}, 216–238}~(2005).

\bibitem{PhysRevResearch.2.033017}
Xiao Chen, Yaodong Li, Matthew P.~A. Fisher, and Andrew Lucas.
\newblock ``Emergent conformal symmetry in nonunitary random dynamics of free
  fermions''.
\newblock \href{https://dx.doi.org/10.1103/PhysRevResearch.2.033017}{Phys. Rev.
  Res. {\bf 2}, 033017}~(2020).

\bibitem{henkel2008non}
M.~Henkel, H.~Hinrichsen, and S.~L\"ubeck.
\newblock ``Non-equilibrium phase transitions: Volume 1: Absorbing phase
  transitions''.
\newblock \href{https://dx.doi.org/10.1007/978-1-4020-8765-3}{Theoretical and
  Mathematical Physics}. Springer Netherlands. ~(2008).

\bibitem{10.21468/SciPostPhysLectNotes.54}
Jacopo Surace and Luca Tagliacozzo.
\newblock ``{Fermionic Gaussian states: an introduction to numerical
  approaches}''.
\newblock \href{https://dx.doi.org/10.21468/SciPostPhysLectNotes.54}{SciPost
  Phys. Lect. NotesPage~54}~(2022).

\bibitem{ravindranath2022robust}
Vikram Ravindranath and Xiao Chen.
\newblock ``Robust oscillations and edge modes in nonunitary floquet systems''.
\newblock \href{https://dx.doi.org/10.1103/PhysRevLett.130.230402}{Phys. Rev.
  Lett. {\bf 130}, 230402}~(2023).

\end{thebibliography}

\end{document}